\documentclass[prc,twocolumn,nofootinbib,showpacs]{revtex4}

\begin{document}

\title{Double, Triple and Hidden Charm Production in 
the Statistical Coalescence Model}

\author{A.P. Kostyuk}

\affiliation{ Institut f\"ur Theoretische Physik and Frankfurt
Institute for Advanced Studies, J.W. Goethe-Universit\"at,
Frankfurt am Main, Germany\\
and \\
Bogolyubov Institute for Theoretical Physics, Kyiv, Ukraine}


\begin{abstract}
The production of particles with  double, triple and 
hidden charm in heavy ion collisions is studied in the framework 
of the statistical coalescence model. According to the postulates 
of the model, the charm quark-antiquark pairs are created at the 
initial stage of a heavy ion reaction in hard parton collisions. 
The amount of charm is assumed to be unchanged at later stages. The 
charm (anti)quarks are distributed among different 
hadron species at hadronization according to the laws of 
statistical physics. Several approaches to the statistical 
treatment of charm hadronization are considered.
The grand canonical approach is appropriate for 
systems containing large number of charm (anti)quarks.
The exact charm conservation and Poissonian 
fluctuations of the number of charm quark-antiquark pairs should be 
taken into account, if the average number of these pairs is of oder of
unity or smaller. The charm hadronization in a subsystem of a 
larger system is discussed. It is explained why the canonical approach is 
not appropriate for the description of charm hadronization.
The obtained formulas can be used to calculate the production of 
charm in heavy ion collisions in a wide energy range. 
\end{abstract}


\maketitle

\section{Introduction}

The thermal hadron gas (HG) model has demonstrated an 
obvious success in describing the chemical composition of 
light-flavored hadrons produced in 
heavy ion \cite{HG} and even in elementary hadron-hadron \cite{HGh} 
as well as in 
electron-positron \cite{HGe} collisions. The experimental data can 
be well fitted with only three free parameters: the temperature 
$T$, the volume $V$ and the baryonic chemical potential $\mu_b$ 
of the hadron gas at the point of the chemical freeze-out.
(Sometimes the fit is improved by introducing one more parameter
--- the strangeness suppression factor $\gamma_s$.)
This success 
motivated attempts to extend the applicability domain 
of the thermal model also to
heavy-flavored hadrons, for instance, to the description
of $J/\psi$ meson production \cite{gago1}. 

A straightforward
application of the equilibrium HG model to hadrons with open and
hidden charm is not, however, justified. Partons with
rather large momenta are needed to produce a heavy quark-antiquark
pair. Consequently, the time of charm equilibration in a
thermal hadronic or even quark-gluon medium is large and
definitely exceeds the lifetime of the fireball. Production of
charm can take place only at the initial stage of the
heavy ion reaction, when hard partons are available.  The
charm production/annihilation rate is too low to keep the
number of heavy quark-antiquark pairs at its chemical equilibrium
value at later stages. Therefore, the {\it total amount} of heavy
flavor should be out of equilibrium at the point of hadronization
and chemical freeze-out. It is reasonable to expect, however, 
the {\it distribution} of heavy quarks and antiquarks among
different hadrons with open and hidden charm to be thermal and
controlled by the same values of thermodynamic parameters that 
fit the chemical composition of light-flavored hadrons. These 
ideas are implemented in the statistical coalescence model (SCM)
\cite{BMS,We,WeBMS}.

In the present paper, I consider SCM, which is based on the 
following postulates: 
\begin{itemize}
\item The charm quarks, $c$, and antiquarks,
$\overline{c}$, are created at the initial stage of A+A reaction
in hard parton collisions.
\item Creation and annihilation of 
$c\overline{c}$ pairs at later stages can be neglected.
\item The formation of observed hadrons with open and hidden
charm takes place near the point of chemical
freeze-out in accordance with the laws of statistical physics.
\end{itemize}
This approach appeared to be quite successful describing the 
$J/\psi$ and $\psi'$ production in (semi)central Pb+Pb collisions at  
Super Proton Synchrotron (SPS) \cite{Kost,psi-prime}. Although the role 
of the statistical coalescence at SPS energies is still under 
discussion \cite{Comb}, its dominance at the energies of the Relativistic 
Heavy Ion Collider (RHIC) and the Large Hadron Collider (LHC) is 
more certain.
The SCM-based predictions for the quarkonium production at RHIC 
\cite{RHIC,RHICa} are consistent with transport calculations 
\cite{brat}.

Quarkonia are not the only type of hadrons whose production 
can be described by the statistical coalescence model. If two, three 
or more charm quark-antiquark pairs are created in a nucleus-nucleus
collision, hadrons with double and triple charm can be formed.
These hadronic states were long time ago predicted by the quark model
(see \cite{Multi} and references therein), but 
most of them have not been observed experimentally yet.\footnote{The recent
observation of the doubly charmed baryon $\Xi_{cc}^{+}$ \cite{Xicc} is 
still under discussion \cite{XiccComm}.}

The hadrons with double and triple charm are of special interest from
both theoretical end experimental point of view. Due to a 
rather large mass of charm (anti)quarks, their interactions within 
a hadron are close to the perturbative regime of Quantum 
Cromodynamics (QCD). A study of the
properties of doubly and especially triply charmed baryons would
allow to test QCD-based models of quark forces. An observation of
more exotic hadronic states, like multi-charm tetra- and pentaquarks
will open
a new window into the structure of the hadronic matter. The intention to
detect the double and triple charm
poses a new challenge to the experimentalists and will demand a further 
development of the experimental technique.

In the present paper, I derive the formulas that allows to calculate 
the yield of double and triple charm particles. The formulas for the 
hidden charm are also included as their detailed derivation has not 
been published yet.

The article is organized as follows. In the Section \ref{Grand} I consider
the grand canonical approach. A system with exactly fixed numbers of
charm quarks and antiquarks is studied in Section \ref{Fixed}.
A more realistic situation, the system with Poissonian fluctuations of 
the number of charm quark-antiquark pairs, is considered in Section
\ref{Poisson}. Section \ref{Subs} is devoted to the case, when only a 
part of the total system is available for the observation: the double,
triple and hidden charm yield in a subsystem is studied. 
Summary is given in Section \ref{Summ}.

\section{The grand canonical approach to the statistical coalescence}
\label{Grand}

The grand canonical version of the statistical coalescence model was
proposed in \cite{BMS}. It can be applied to the systems containing large 
($N_{c\overline{c}} \gg 1$) number of $c\overline{c}$ pairs.

Let us start from the grand canonical partition function for the ideal 
hadron gas in the Boltzmann approximation:
\begin{equation}\label{gcpf}
{\cal Z}(V,T,\{ \lambda \} ) = \sum_{i_1 = 0}^{\infty} \sum_{i_2 = 0}^{\infty}
\dots \sum_{i_L = 1}^{\infty} \prod_{l=0}^L
\frac{\left[ \lambda_l \phi(T,m_l,g_l) V \right]^{i_l}}{i_l !},
\end{equation}
where $L$ is the total number of hadron species (including resonances), 
$V$ and $T$ are the volume and temperature, 
$\phi(T;m,g)$ is the one-particle partition function
\begin{eqnarray}\label{phi}
\phi(T;m,g) &=& \frac{g}{2 \pi^2} \int_0^{\infty}p^2 dp~
\exp \left( - \frac{\sqrt{p^{2}+m^{2}}}{T}  \right)~ \\
&=& g\frac{m^{2} T}{2 \pi^{2}}~ K_{2}\left( \frac{m}{T} \right)~.
\nonumber
\end{eqnarray}
Here $m$ is the particle mass, $g$ is the degeneration
factor (the number of spin states), $K_{2}$ is the modified Bessel 
function.
In the nonrelativistic limit $m >> T$ the expression (\ref{phi}) takes
the form
\begin{equation}\label{phinr}
\phi(T;m,g)~ \simeq~  g \left(\frac{mT}{2 \pi} \right)^{3/2} \exp(-m/T).
\end{equation}

The fugacity $\lambda_l$ is expressed via the chemical potentials 
(electric --- $\mu_e$, baryonic --- $\mu_b$, strangeness --- $\mu_s$ and 
charm --- $\mu_c$)
and suppression (enhancement) factors 
(for strangeness --- $\gamma_s$, and charm --- $\gamma_c$):
\begin{eqnarray}\label{lambda}
\lambda_l &=& \gamma_s^{|s|_l} \gamma_c^{|c|_l}
\exp\left( \frac{\mu_l}{T} \right),  \\\label{mu}
\mu_l &=& q_l \mu_e + b_l \mu_b + s_l \mu_s + c_l \mu_c \ , \label{mul}
\end{eqnarray}
where $q_l$, $b_l$, $s_l$, $c_l$,  are, respectively, the electric charge, 
baryon number, strangeness and charm of the hadron species $l$, 
$|s|_l$ and $|c|_l$ are the numbers of valence strange and charmed 
(anti-)quarks. The chemical potentials in the right-hand-side of 
(\ref{mul}) are responsible for keeping
the correct {\it average} values of the corresponding charges in the system.
The suppression (enhancement) factors $\gamma_s$ and $\gamma_c$
are introduced to take into account a 
deviation of the total number of strangeness and charm from their equilibrium 
values\footnote{The suppression (enhancement) factor $\gamma_s$ 
(the same for $\gamma_c$)
is equivalent to an additional chemical potential $\mu_{|s|}$ 
which, in contrast to $\mu_{s}$,  has the same influence on the 
strangeness as on the antistrangeness: 
$\gamma_s = \exp\left( \frac{\mu_{|s|}}{T} \right)$}. 

The average number of particles is given for each species as 
\begin{equation}\label{Nl}
N_l = \lambda_l
\frac{\partial \log {\cal Z}(V,T,\{ \lambda \} )}
{\partial \lambda_l} = \lambda_l \phi(T,m_l,g_l) V.
\end{equation}

The total number of hadrons 
\begin{equation}\label{Nt}
N_{tot} = \sum_l N_l, \mbox{  ($l$ runs over all hadron species)},
\end{equation}
can be broken up into several pieces:
$N_0$, the number of zero charm hadrons (excluding hidden charm),
$N_H$, the number of hidden charm mesons,
$N_1$, $N_{\bar{1}}$, $N_2$, $N_{\bar{2}}$, $N_3$, $N_{\bar{3}}$, the
numbers of hadrons with, respectively, single, double and triple 
charm and anticharm:
\begin{equation}\label{Ntc}
N_{tot} = N_0+N_H+N_1+N_{\bar{1}}+N_2+N_{\bar{2}}+N_3+N_{\bar{3}} .
\end{equation}

Let us consider a system containing {\it in average} $N_{c}$ charm quark and 
$N_{\bar{c}}$ charm antiquarks. Then, as far as charm creation and annihilation
are neglected, the following equalities should be satisfied:
\begin{eqnarray}
\label{gceq1} \langle N_{c} \rangle &=& N_1 + N_H + 2 N_2 + 3 N_3,\\ 
\label{gceq2} \langle N_{\bar{c}} \rangle &=& N_{\bar{1}} + N_H + 
2 N_{\bar{2}} + 3 N_{\bar{3}}.
\end{eqnarray}
It is easy to see from (\ref{lambda}),(\ref{mu}) and (\ref{Nl}) that the 
above equations can be rewritten as 
\begin{eqnarray}
\label{gceqt1} \langle N_{c} \rangle &=& \gamma_c \lambda_c \tilde{N}_1 + \gamma_c^2
\tilde{N}_H + 2 \gamma_c^2 \lambda_c^2 \tilde{N}_2 + 3 \gamma_c^3
\lambda_c^3 \tilde{N}_3, \label{eq1} \\ 
\label{gceqt2} \langle N_{\bar{c}}  \rangle &=& \gamma_c
\lambda_c^{-1} \tilde{N}_{\bar{1}} + \gamma_c^2 \tilde{N}_H + 2
\gamma_c^2 \lambda_c^{-2} \tilde{N}_{\bar{2}} + 3 \gamma_c^3
\lambda_c^{-3} \tilde{N}_{\bar{3}}, \nonumber \\
\label{eq2}
\end{eqnarray}
where 
\begin{eqnarray}\label{Ntilde}
\tilde{N}_k &=& \left. N_k \right|_{\mu_c=0, \gamma_c=1},\ \ 
k= H, 1, \bar{1}, 2, \bar{2}, 3, \bar{3},
\end{eqnarray}
and 
\begin{eqnarray}
\lambda_c = \exp\left( \frac{\mu_c}{T} \right).
\end{eqnarray}

The constituent charm quark mass is about an order of magnitude 
larger than the typical temperature of chemical freeze-out.
From this reason,
\begin{equation}\label{inequ}
\tilde{N}_1,\tilde{N}_{\bar{1}} \gg \tilde{N}_H,
\tilde{N}_2,\tilde{N}_{\bar{2}} \gg
\tilde{N}_3,\tilde{N}_{\bar{3}} 
\end{equation}
due to the exponential factor in (\ref{phinr}). Therefore, 
if the factor $\gamma_c$ is not extraordinary large,
the terms in (\ref{gceqt1}) and (\ref{gceqt2})
corresponding to hidden, double and triple (anti)charm can be neglected 
in a zero approximation and the coupled equations (\ref{gceqt1}) and 
(\ref{gceqt2}) can be simplified:
\begin{eqnarray}
\label{gceqt10} \langle N_{c}  \rangle  &=& \gamma_c^{(0)} \lambda_c^{(0)} \tilde{N}_1 ,
\label{eq1o} \\ 
\label{gceqt20}  \langle N_{\bar{c}}  \rangle &=&
\gamma_c^{(0)} (\lambda_c^{(0)})^{-1} \tilde{N}_{\bar{1}} \label{eq2o}.
\end{eqnarray}
The solution can be easily found:
\begin{eqnarray}
\gamma_c^{(0)} &=& \sqrt{\frac{ \langle N_{c} \rangle  
\langle N_{\bar{c}}  \rangle }{\tilde{N}_1
\tilde{N}_{\bar{1}}}},\label{gamcG}\\ 
\lambda_c^{(0)} &=& \sqrt{  \frac{ \langle N_{c} \rangle
\tilde{N}_{\bar{1}} }{\langle N_{\bar{c}} \rangle \tilde{N}_1}}.
\label{lamcG}
\end{eqnarray}
Now one can calculate the number of single, hidden, double and  triple 
charm particles in the zero approximation:
\begin{eqnarray}
\langle N_1 \rangle_{GCE}^{(0)} &=& \gamma_c^{(0)} \lambda_c^{(0)}
\tilde{N}_1 = \langle N_{c}  \rangle ,  \label{N1G}\\
\langle N_H \rangle_{GCE}^{(0)} &=& (\gamma_c^{(0)})^2
\tilde{N}_H = \langle N_{c} \rangle  
\langle N_{\bar{c}} \rangle 
\frac{\tilde{N}_H}{\tilde{N}_1 \tilde{N}_{\bar{1}}}, \label{NHG} \\
\langle N_2 \rangle_{GCE}^{(0)} &=& (\gamma_c^{(0)})^2 (\lambda_c^{(0)})^2
\tilde{N}_2 = \langle N_{c} \rangle^2
\frac{\tilde{N}_2}{\tilde{N}_1^2}, \label{N2G} \\ 
\langle N_3 \rangle_{GCE}^{(0)} &=& (\gamma_c^{(0)})^2 (\lambda_c^{(0)})^2 
\tilde{N}_3 =
\langle N_{c} \rangle^3 \frac{\tilde{N}_3}{\tilde{N}_1^3}. \label{N3G} 
\end{eqnarray}
The formulas for anticharm can be obtained by the obvious replacement 
$c \rightarrow \bar{c}$, $1 \rightarrow \bar{1}$ etc.

The formulas (\ref{N1G}--\ref{N3G}) are obtained under the assumption
that the charm enhancement factor $\gamma_s$ is not large, so 
that only a tiny
fraction of the charm quarks and antiquarks hadronizes into hidden,
double and triple charm particles. This may be not true  
at very high energies. Indeed, let us consider for example the leading 
order expression (\ref{N2G}) for the number of double charm particles.
The fraction of charm quarks that hadronize 
into double charm particles is proportional to the number of 
charm quarks in the system and inversely proportional to the system
volume at freeze-out:
\begin{equation}\label{ratio2}
\frac{\langle N_2 \rangle_{GCE}^{(0)}}{\langle N_{c} \rangle}
\propto \frac{\langle N_{c} \rangle}{V}.
\end{equation}
The volume is proportional to the multiplicity of light hadrons.
The charm production cross section grows
with the collision energy faster than the multiplicity of light hadrons.
Therefore, the ratio 
(\ref{ratio2}) becomes comparable to unity at some point. In this case the 
above approximation does not work.

The accuracy of the leading-order approximation 
(\ref{N1G})--(\ref{N3G}) can be 
estimated by calculating the next-to-leading order corrections. 
Let us substitute $\gamma_c \rightarrow \gamma_c^{(0)} + \gamma_c^{(1)}$ and 
$\lambda_c \rightarrow \lambda_c^{(0)} + \lambda_c^{(1)}$ into 
(\ref{eq1}) and (\ref{eq2}). Neglecting higher order terms like 
$\gamma_c^{(1)} \lambda_c^{(1)}$, $\gamma_c^{(1)} \tilde{N}_k$
($k= H, 2, \bar{2}, 3, \bar{3}$), etc. and taking into account 
(\ref{eq1o}),(\ref{eq2o}) and (\ref{NHG})--(\ref{N3G}) 
one gets
\begin{widetext}
\begin{eqnarray}
\lambda_c^{(0)} \gamma_c^{(1)} + \gamma_c^{(0)} \lambda_c^{(1)} &=&
- \frac{\langle N_H \rangle_{GCE}^{(0)} + 
2 \langle N_2 \rangle_{GCE}^{(0)} + 
3 \langle N_3 \rangle_{GCE}^{(0)}}{\tilde{N}_1} , \label{eq1o1}\\
\left( \lambda_c^{(0)} \right)^{-2} \left(
\lambda_c^{(0)} \gamma_c^{(1)} - \gamma_c^{(0)} \lambda_c^{(1)}
\right)
&=&
- \frac{\langle N_H \rangle_{GCE}^{(0)} + 
2 \langle N_{\bar{2}} \rangle_{GCE}^{(0)} + 
3 \langle N_{\bar{3}} \rangle_{GCE}^{(0)}}{\tilde{N}_{\bar{1}}}. 
\label{eq2o1}
\end{eqnarray}
These coupled linear equations can be easily solved:
\begin{eqnarray}
\gamma_c^{(1)}  &=& - \frac{1}{2} \gamma_c^{(0)}
\left(
\frac{\langle N_H \rangle_{GCE}^{(0)} + 
2 \langle N_2 \rangle_{GCE}^{(0)} + 
3 \langle N_3 \rangle_{GCE}^{(0)}}{\langle N_{c} \rangle} 
+
\frac{\langle N_H \rangle_{GCE}^{(0)} + 
2 \langle N_{\bar{2}} \rangle_{GCE}^{(0)} + 
3 \langle N_{\bar{3}} \rangle_{GCE}^{(0)}}{\langle N_{\bar{c}} \rangle}
\right), \\
\lambda_c^{(1)}  &=& - \frac{1}{2} \lambda_c^{(0)}
\left(
\frac{\langle N_H \rangle_{GCE}^{(0)} + 
2 \langle N_2 \rangle_{GCE}^{(0)} + 
3 \langle N_3 \rangle_{GCE}^{(0)}}{\langle N_{c} \rangle} 
-
\frac{\langle N_H \rangle_{GCE}^{(0)} + 
2 \langle N_{\bar{2}} \rangle_{GCE}^{(0)} + 
3 \langle N_{\bar{3}} \rangle_{GCE}^{(0)}}{\langle N_{\bar{c}} \rangle}
\right).
\end{eqnarray}
\end{widetext}
Now the next-to-leading order corrections to the number of particles with
different charm content can be calculated:
\begin{eqnarray}
\langle N_1 \rangle_{GCE}^{(1)} &=& - \left(
\langle N_H \rangle_{GCE}^{(0)} + 
2 \langle N_2 \rangle_{GCE}^{(0)} + 
3 \langle N_3 \rangle_{GCE}^{(0)} \right), \nonumber \\ 
\label{N1Go1} \\
\langle N_H \rangle_{GCE}^{(1)} &=&  \langle N_H \rangle_{GCE}^{(0)}
\left(
\frac{\langle N_1 \rangle_{GCE}^{(1)}}{\langle N_{c} \rangle} + 
\frac{\langle N_{\bar{1}} \rangle_{GCE}^{(1)}}
{\langle N_{\bar{c}} \rangle} 
\right), \label{NHGo1} \\
\langle N_2 \rangle_{GCE}^{(1)} &=& \langle N_2 \rangle_{GCE}^{(0)}  
\frac{2 \langle N_1 \rangle_{GCE}^{(1)}}{\langle N_{c} \rangle} ,
\label{N2Go1} \\
\langle N_3 \rangle_{GCE}^{(1)} &=& \langle N_3 \rangle_{GCE}^{(0)}  
\frac{3 \langle N_1 \rangle_{GCE}^{(1)}}{\langle N_{c} \rangle} .
\label{N3Go1}
\end{eqnarray}
The higher order corrections can be obtained in a similar way. Still, 
the expressions become rather unwieldy. It is more reasonable to solve 
the coupled equations (\ref{eq1}) and (\ref{eq2}) numerically, if the 
next-to-leading order corrections (\ref{N1Go1})--(\ref{N3Go1}) become 
too large.

The above formulas allow to calculate the total number of particles with 
given (anti)charm content (hidden, single, double or triple). If the 
particle number of a single species is needed, it can be found in the 
following way. One can 
deduce from equations (\ref{lambda}), (\ref{mu}) and (\ref{Nl}) that the 
number of particles of a
single species $l$ can be found from
\begin{equation}\label{Nlc}
N_{l} = \frac{\tilde{N}_l}{\tilde{N}_k} N_{k},
\end{equation}
where $k= H, 1, \bar{1}, 2, \bar{2}, 3$ or $\bar{3}$, depending on the 
charm content of the species $l$.

The grand canonical approach is the simplest version of the statistical
coalescence model. However, it does not reflect the real experimental 
situation sufficiently well. Indeed,  we operated only with average values 
$\langle N_{c} \rangle$ and  $\langle N_{\bar{c}} \rangle$ in the above 
consideration. The correlations between $N_{c}$ and  
$N_{\bar{c}}$ were ignored. In reality, however, the quarks and antiquarks
are produced in pairs. So that not only average numbers coincide
$\langle N_{c} \rangle = \langle N_{\bar{c}} \rangle$, but also the numbers
of quarks and antiquarks {\it in every single event} are equal: 
$N_{c} = N_{\bar{c}}$. The influence of this fact on the results of the 
SCM will be studied in the subsequent sections.

\section{System with fixed numbers of charm quarks and antiquarks}
\label{Fixed}

In this section, a system with fixed numbers of charm quarks 
and antiquarks is considered. This does not correspond exactly
to what is observed in the experiment. Still a study of such a system 
would allow us  to obtain an important intermediate 
result. Later, it will be used in more realistic calculations.

Using Eq.(\ref{Nl}), one can rewrite the partition function (\ref{gcpf})
in a more compact form:
\begin{eqnarray}\label{gcpfc}
{\cal Z}(V,T,\{ \lambda \} ) &=& \sum_{i_1 = 0}^{\infty} \sum_{i_2 = 0}^{\infty}
\dots \sum_{i_L = 1}^{\infty} \prod_{l=0}^L
\frac{N_l^{i_l}}{i_l !} \nonumber \\
&=& \exp \left( \sum_{l=0}^L N_l \right)  = \exp \left( N_{tot} \right) .
\end{eqnarray}

The total number of particles in the last expression  
can be represented according to (\ref{Ntc}), then
\begin{eqnarray}\label{gcpfc1}
{\cal Z}(V,T,\{ \lambda \} ) &=& \exp \left( \sum_{k} N_k \right)
= \prod_{k} \exp \left( N_k \right) , \\ 
& & k= 0, H, 1, \bar{1}, 2, \bar{2}, 3, \bar{3}. \nonumber
\end{eqnarray}
Expanding all the exponents in the product, except the first one, into 
the Taylor series one gets 
\begin{eqnarray}\label{gcpf1}
{\cal Z}(V,T,\{ \lambda \} )  &=&
e^{N_0} \sum_{i_H = 0}^{\infty} \sum_{i_1 = 0}^{\infty}
\dots \sum_{i_{\bar{3}} = 1}^{\infty} \prod_{k}
\frac{N_k^{i_k}}{i_k !}, \\ 
& & k= H, 1, \bar{1}, 2, \bar{2}, 3, \bar{3}. \nonumber
\end{eqnarray}

Due to the property of The Kronecker delta $\sum_m \delta(m,n) = 1$,
nothing changes if we multiply an expression by a Kronecker delta 
and sum over one of its indices. Doing this twice on the expression 
(\ref{gcpf1}), one gets
\begin{widetext}
\begin{eqnarray}\label{gcpf1a}
{\cal Z}(V,T,\{ \lambda \} )  &=&
e^{N_0} \sum_{i_H = 0}^{\infty} \sum_{i_1 = 0}^{\infty}
\dots \sum_{i_{\bar{3}} = 1}^{\infty} 
\sum_{N_{c}} \sum_{N_{\bar{c}}}
\delta(N_{c},i_H+i_1+i_2+i_3) 
\delta(N_{\bar{c}},i_H+i_{\bar{1}}+i_{\bar{2}}+i_{\bar{3}})
\prod_{k}
\frac{N_k^{i_k}}{i_k !}, \\ 
& & k= H, 1, \bar{1}, 2, \bar{2}, 3, \bar{3}. \nonumber
\end{eqnarray}
Then, after changing the summation order and using (\ref{lambda}), 
(\ref{mul}), (\ref{Nl}) and (\ref{Ntilde}), the above expression can be 
rewritten as 
\begin{eqnarray}\label{gcpf2}
{\cal Z}(V,T,\{ \lambda \} ) &=& e^{N_0} \sum_{N_{c}} \sum_{N_{\bar{c}}}
\gamma_c^{N_{c} + N_{\bar{c}}} 
\exp\left[ (N_{c} - N_{\bar{c}}) \frac{\mu_c}{T} \right]
Z_{N_{c} N_{\bar{c}}}(V,T,\{ \tilde{\lambda} \}) .
\end{eqnarray}
Here
$Z_{N_{c} N_{\bar{c}}}$ is the partition functions for the systems
containing exactly  $N_{c}$ charm quarks and $N_{\bar{c}}$ antiquarks:
\begin{eqnarray}\label{cpf}
Z_{N_{c} N_{\bar{c}}}(V,T,\{ \tilde{\lambda} \})
&=&  \sum_{i_H = 0}^{\infty} \sum_{i_1 = 0}^{\infty}
\dots \sum_{i_{\bar{3}} = 1}^{\infty}  
\delta(N_{c},i_H+i_1+i_2+i_3) \,
\delta(N_{\bar{c}},i_H+i_{\bar{1}}+i_{\bar{2}}+i_{\bar{3}})
\prod_{k}
\frac{\tilde{N}_k^{i_k}}{i_k !},\\
& & \hspace{5cm} k= H, 1, \bar{1}, 2, \bar{2}, 3, \bar{3}. \nonumber
\end{eqnarray}
This function is {\it canonical} with respect to the {\it exact} 
conservation of the number of charm quarks and antiquarks and 
{\it grand canonical} with respect to the conservation {\it in average} 
of all other charges, whose values are controlled by the activities 
\begin{equation}
\tilde{\lambda}_k = \left. \lambda_k \right|_{\mu_c=0, \gamma_c=1} .
\end{equation}

The number of summations in (\ref{cpf}) can be reduced due to 
the Kronecker deltas:
\begin{equation}\label{gcpf2a}
Z_{N_{c} N_{\bar{c}}}(V,T,\{ \tilde{\lambda} \})
=
\sum_{i_H = 0}^{i_H^{max}} \frac{\tilde{N}_H^{i_H}}{i_H !}
\sum_{i_2 = 0}^{i_2^{max}} \frac{\tilde{N}_2^{i_2}}{i_2 !}
\sum_{i_{\bar{2}} = 0}^{i_{\bar{2}}^{max}}
\frac{\tilde{N}_{\bar{2}}^{i_{\bar{2}}}}{i_{\bar{2}} !}
\sum_{i_3 = 0}^{i_3^{max}} \frac{\tilde{N}_3^{i_3}}{i_3 !}
\sum_{i_{\bar{3}} = 0}^{i_{\bar{3}}^{max}}
\frac{\tilde{N}_{\bar{3}}^{i_{\bar{3}}}}{i_{\bar{3}} !}
\frac{\tilde{N}_1^{i_1}}{i_1 !}
\frac{\tilde{N}_{\bar{1}}^{i_{\bar{1}}}}{i_{\bar{1}} !}.
\end{equation}
\end{widetext}
Here 
\begin{eqnarray}
i_H^{max} &=& min(N_{c},N_{\bar{c}}),\\
i_2^{max}(i_H)  &=& [(N_{c}-i_H)/2],\\
i_{\bar{2}}^{max}(i_H)  &=& [(N_{\bar{c}}-i_H)/2],\\
i_3^{max}(i_H,i_2) &=& [(N_{c}-i_H-2 i_2)/3],\\
i_{\bar{3}}^{max}(i_H,i_{\bar{2}}) &=& [(N_{\bar{c}}-i_H - 2 i_{\bar{2}})/3],\\
i_1(i_H,i_2,i_3) &=& [N_{c} - i_H - 2 i_2 - 3 i_3],\\
i_{\bar{1}}(i_H,i_{\bar{2}},i_{\bar{3}}) &=&
[N_{\bar{c}}-i_H - 2 i_{\bar{2}}- 3 i_{\bar{3}}].
\end{eqnarray}
(The square brackets mean here the integer part, i.e. $[x]$ is the largest
integer number that does not exceed $x$.)

The average number of hidden (single, double, triple) (anti)charm particles 
in the system with {\it fixed} numbers of charm quarks and antiquarks is 
found as 
\begin{eqnarray}\label{Nk}
& & 
\langle N_{k} \rangle_{fix}  =  \sum_l \tilde{\lambda}_l
\frac{\partial \log Z_{N_{c}N_{\bar{c}}} }{\partial \tilde{\lambda}_l} =
\sum_l \tilde{\lambda}_l \frac{\partial \tilde{N}_k}{\partial
\tilde{\lambda}_l} \frac{\partial \log Z_{N_{c}N_{\bar{c}}}}{\partial
\tilde{N}_k}, \nonumber \\
& & k= H, 1, \bar{1}, 2, \bar{2}, 3, \bar{3}; \\
& & \mbox{$l$ runs over all hadron species of the type $k$}.
\nonumber
\end{eqnarray}
From (\ref{Nl}) one sees that 
\begin{equation}
\tilde{\lambda}_l \frac{\partial \tilde{N}_l}{\partial
\tilde{\lambda}_l} =  \tilde{N}_l,
\end{equation}
therefore 
\begin{equation}
\sum_l \tilde{\lambda}_l 
\frac{\partial \tilde{N}_k}{\partial \tilde{\lambda}_l}
= \tilde{N}_k
\end{equation}
and finally
\begin{equation}\label{Nkf}
\langle N_{k} \rangle_{fix} = \tilde{N}_k
\frac{1}{Z_{N_{c}N_{\bar{c}}}}
\frac{\partial Z_{N_{c}N_{\bar{c}}}}{\partial \tilde{N}_k}.
\end{equation}

I restrict my further consideration to the case, when the 
partition function (\ref{gcpf2a}) is dominated\footnote{Due 
to the inequality (\ref{inequ}), the contribution of 
other terms becomes sizable only at very large $N_{c}$ and/or 
$N_{\bar{c}}$. This case can be treated within the grand canonical 
approach and there is no reason to study it here.} by the 
term with
$i_H = i_2= i_{\bar{2}} = i_3 = i_{\bar{3}} = 0$, 
$i_1 = N_{c}$ and  $i_{\bar{1}} = N_{\bar{c}}$:
\begin{equation}\label{Z0}
Z_{N_{c}N_{\bar{c}}}^{(0)} \approx 
\frac{\tilde{N}_1^{N_{c}}}{N_{c} !}
\frac{\tilde{N}_{\bar{1}}^{N_{\bar{c}}}}{N_{\bar{c}}!} .
\end{equation} 
The same term dominates also the derivatives of 
$Z_{N_{c}N_{\bar{c}}}$ with respect to 
$N_{c}$ and $N_{\bar{c}}$. It is easy to see that in this case, most
of the charm hadronizes into hadrons containing 
only one $c$-quark or antiquark:
\begin{eqnarray}
\langle N_{1} \rangle_{fix}^{(0)}  &\approx& N_{c}, \label{N1f} \\
\langle N_{\bar{1}} \rangle_{fix}^{(0)}  
&\approx& N_{\bar{c}}. \label{N1bf}
\end{eqnarray}
Only a tiny fraction of the total charm is accommodated into  hidden, 
double and triple charm hadrons.

The leading term (\ref{Z0}) does not depend on $i_H$. Therefore, 
the derivative ${\partial Z_{N_{c}N_{\bar{c}}}}/{\partial \tilde{N}_H}$
is dominated by the term of (\ref{gcpf2a}) with  $i_H = 1$,
$i_2= i_{\bar{2}} = i_3 = i_{\bar{3}} = 0$, 
$i_1 = N_{c}-1$ and  $i_{\bar{1}} = N_{\bar{c}}-1$:
\begin{equation}\label{dZH}
\frac{\partial Z_{N_{c}N_{\bar{c}}}^{(0)}}{\partial \tilde{N}_H}
\approx 
\frac{\tilde{N}_1^{N_{c}-1}}{(N_{c}-1) !}
\frac{\tilde{N}_{\bar{1}}^{N_{\bar{c}}-1}}{(N_{\bar{c}}-1)!} .
\end{equation}
From (\ref{Nkf}) one finds the average number of hidden charm particles 
in the system with fixed numbers of charm quarks and antiquarks:
\begin{equation}\label{NHf}
\langle N_{H} \rangle_{fix}^{(0)} = 
N_{c} N_{\bar{c}} \frac{\tilde{N}_H}{\tilde{N}_1 \tilde{N}_{\bar{1}}} .
\end{equation}

Similarly, the average numbers of double and triple charm particles are 
given by:
\begin{equation}\label{N2f}
\langle N_{2} \rangle_{fix}^{(0)} = 
N_{c} (N_{c} - 1) \frac{\tilde{N}_2}{\tilde{N}_1^2} 
\end{equation}
and 
\begin{equation}\label{N3f}
\langle N_{3} \rangle_{fix}^{(0)} = 
N_{c} (N_{c} - 1) (N_{c} - 2) \frac{\tilde{N}_3}{\tilde{N}_1^3},
\end{equation}
respectively. The corresponding formulas for anticharm can be obtained 
from (\ref{N2f}) and (\ref{N3f}) by the obvious replacement 
$c \rightarrow \bar{c}$, $1 \rightarrow \bar{1}$ etc.

We have considered the charm hadron production by quark coalescence in 
the thermal hadron system with {\it fixed} numbers of charm quarks and 
antiquarks. In reality, this number is not fixed. It
fluctuates from event to event. These fluctuations have to be taken into 
account in more realistic calculations.

\section{The system with Poissonian fluctuations of the number of charm
quark-antiquark pairs}
\label{Poisson}

In a relativistic collision of two nuclei, the charm quarks 
and antiquarks are created in pairs in independent nucleon-nucleon 
collisions. Therefore, the number 
of quarks in the system is always equal to the number of antiquarks:
\begin{equation}\label{Nccbar}
N_{c} = N_{\bar{c}} \equiv N_{c\bar{c}}
\end{equation}
and the fluctuations of the number of pairs 
approximately\footnote{This approximation obviously breaks down 
at very large $N_{c\bar{c}}$ when the energy of the produced 
$c\bar{c}$ pairs becomes comparable with the total energy of the system.
But the probability of such events is clearly negligible. Only a tiny fraction
of the total energy of the system is accumulated into the charm particles.}
conforms  the Poissonian law:
\begin{equation}\label{prob} 
w_{P}(N_{c\bar{c}}) = e^{- \langle N_{c\bar{c}} \rangle } 
\frac{\langle N_{c\bar{c}} \rangle^{N_{c\bar{c}}}}{N_{c\bar{c}}!}.
\end{equation}
Here $w(N_{c\bar{c}})$ is the probability to observe $N_{c\bar{c}}$ charm 
quark-antiquark pairs in an event, provided that the average number of 
$c\bar{c}$ pairs in this type of events is $\langle N_{c\bar{c}} \rangle$.

The average number of particles with hidden, single, double, and triple
charm is found by the convolution of the results of the last 
section (\ref{N1f}),(\ref{NHf})--(\ref{N3f}) with 
the probability (\ref{prob}):
\begin{eqnarray}\label{NkP}
\langle N_{k} \rangle_{P} &=& \sum_{N_{c\bar{c}}=1}^{\infty} 
w_{P}(N_{c\bar{c}})
\langle N_{k} \rangle_{fix} \\
& & k= H, 1, \bar{1}, 2, \bar{2}, 3, \bar{3} \nonumber
\end{eqnarray}
(The subscript ``$P$'' stands for ``Poisson'').
Here we again see that the most of the charm is accommodated into the single
charm hadrons:
\begin{equation} \label{N1P}
\langle N_{1} \rangle_{P}^{(0)}  \approx
\langle N_{\bar{1}} \rangle_{P}^{(0)}  \approx
\langle N_{c\bar{c}} \rangle. 
\end{equation}
The rest (a tiny fraction) is distributed over hidden, double and triple charm 
particles whose number can be easily found:
\begin{eqnarray}
\langle N_H \rangle_{P}^{(0)} &=&  \langle N_{c\bar{c}} \rangle  
(\langle N_{c\bar{c}} \rangle +1) 
\frac{\tilde{N}_H}{\tilde{N}_1 \tilde{N}_{\bar{1}}}, \label{NHP} \\
\langle N_2 \rangle_{P}^{(0)} &=&  \langle N_{c\bar{c}} \rangle^2
\frac{\tilde{N}_2}{\tilde{N}_1^2}, \label{N2P}\\ 
\langle N_3 \rangle_{P}^{(0)} &=& 
\langle N_{c\bar{c}} \rangle^3 \frac{\tilde{N}_3}{\tilde{N}_1^3}. \label{N3P}
\end{eqnarray}
Surprisingly, the formulas for the number of double and triple 
charm particles (\ref{N2P}),(\ref{N3P}) appeared to be exactly the same 
as in the 
grand canonical approach (\ref{N2G}),(\ref{N3G}). 

However, this is not the case for the hidden charm. If the average 
number of $c\bar{c}$ pairs is small, 
$\langle N_{c\bar{c}} \rangle \alt 1$, the average number of produced 
hidden charm hadrons 
(\ref{NHP}) is essentially larger than it would be na\"\i vely 
expected from the  grand canonical formula (\ref{NHG}). The two formulas 
(\ref{NHG}) and (\ref{NHP})
give similar results only when the number $c\bar{c}$ pairs is large,
$\langle N_{c\bar{c}} \rangle \gg 1$. The reason for this similarity is that 
the Poissonian distribution becomes narrow at 
$\langle N_{c\bar{c}} \rangle \gg 1$:
$\langle (N_{c\bar{c}} - \langle N_{c\bar{c}} \rangle)^2 \rangle =
\langle N_{c\bar{c}} \rangle
\ll \langle N_{c\bar{c}} \rangle^2$.

It is easy to see that any narrow probability distribution  
of the number of charm quarks and antiquarks would give
the same result as the grand canonical approach, provided that 
$\langle N_{c\bar{c}} \rangle \gg 1$.

\section{Charm coalescence in a subsystem}
\label{Subs}

The formulas of the previous section allow to calculate the 
number of charm particles in the entire system. It often happens, however, 
that only a part of the phase space is observed in the experiment: a limited
rapidity interval, for instance. Formulas for charm production in a 
subsystem are necessary in this case. The charm and non-charm hadrons are 
distributed inhomogeneously in the (phase)
space.  Therefore, the number of observed 
charm particles in a subsystem cannot be calculated merely 
as a fraction of their total number in the system, proportional to 
the ratio of the volume of the subsystem to the total volume of the 
system. Moreover, the total volume of the system 
may be even unknown. The formulas derived in this section are based on 
the thermodynamic parameters of the subsystem without any reference to those 
of the entire system.

Let $\xi \le 1$ is the probability to find a charm quark in
the subsystem, provided that exactly one $c\bar{c}$ pair is present 
in the entire system. Then, if the number of $c\bar{c}$ pairs in the 
entire system is $N_{c\bar{c}}$, the probability to find $N_{c}$ charm quarks
in the subsystem is given by the binomial law:
\begin{equation}\label{binom}
w(N_{c} | N_{c\bar{c}}) = 
\frac{N_{c\bar{c}}!}{N_{c}! (N_{c\bar{c}} - N_{c})!} 
\xi^{N_{c}} (1-\xi)^{N_{c\bar{c}}-N_{c}}.
\end{equation}
It is assumed that the distributions of quarks and antiquarks are 
uncorrelated, and the probability distribution of the number
of antiquarks $N_{\bar{c}}$ is given by the same binomial law. 

Then, if the total number of $c\bar{c}$ pairs in the system is fixed,
the average number of the hidden charm particles produced in the system 
is given by a convolution of two probability distributions 
(\ref{binom}) with the right-hand side of (\ref{NHf}): 
\begin{eqnarray}\label{NHfs}
\langle N_{H} \rangle_{fix}^{(0)} &=& 
\sum_{N_{c}=0}^{\infty} w(N_{c} | N_{c\bar{c}})
\sum_{N_{c}=0}^{\infty} w(N_{\bar{c}} | N_{c\bar{c}})
N_{c} N_{\bar{c}}
\frac{\tilde{N}_H}{\tilde{N}_1 \tilde{N}_{\bar{1}}}
\nonumber \\
&=& 
\xi^2
(N_{c\bar{c}})^2 \frac{\tilde{N}_H}{\tilde{N}_1 \tilde{N}_{\bar{1}}} .
\end{eqnarray}

In a more realistic situation, if the total number of $c\bar{c}$ pairs 
in the entire system fluctuates according to the Poissonian law (\ref{prob}), 
the average number of hidden charm in the subsystem is given by 
\begin{equation}
\langle N_H \rangle_{sub}^{(0)} = \xi^2  \langle N_{c\bar{c}} \rangle  
(\langle N_{c\bar{c}} \rangle +1) 
\frac{\tilde{N}_H}{\tilde{N}_1 \tilde{N}_{\bar{1}}}. \label{NHPs}
\end{equation}
Similarly for the double and triple charm:
\begin{eqnarray}
\langle N_2 \rangle_{sub}^{(0)} &=&  \xi^2 \langle N_{c\bar{c}} \rangle^2
\frac{\tilde{N}_2}{\tilde{N}_1^2} , \label{N2Ps}\\ 
\langle N_3 \rangle_{sub}^{(0)} &=& \xi^3
\langle N_{c\bar{c}} \rangle^3 \frac{\tilde{N}_3}{\tilde{N}_1^3}. \label{N3Ps}
\end{eqnarray}
Note, that  only the average number of the charm quark-antiquark pairs 
refers to the entire system in the above equations. The thermal 
quantities ${N}_H$, $\tilde{N}_1$ and $\tilde{N}_{\bar{1}}$ are related 
to the subsystem under consideration. Therefore, one can calculate the number 
of charm particles produced in the subsystem, even if the thermodynamic 
parameters of the entire system are not known. Moreover, to calculate 
the number of double and triple charm, one does not actually need the 
total number of $c\bar{c}$ pairs in the entire system. It suffices to know 
this number for the subsystem $\xi \langle N_{c\bar{c}} \rangle$. Only the 
calculation of the hidden charm requires the total number of charm pairs 
in the entire system, if this number is comparable to $1$ or smaller.

\section{The canonical approach and why it is not appropriate for 
the charm coalescence}
\label{Canon}

The canonical ensemble (CE) approach was initially proposed for a 
thermal treatment of the strangeness productions, when the average 
number of the strange particles in the system is small ($\alt 1$)
and the exact conservation of the net strangeness becomes important  
\cite{CanonStr}. It was also applied to the baryonic charge \cite{CanonBar}.

Although, as it will be explained later, this approach is not appropriate 
for the charm coalescence, I consider it in details because of two
reasons. First, it has been widely used for the description of the charmonium 
production \cite{We,WeBMS,RHICa,Rep}. Second, it would be instructive to see, 
what changes, if the event-by-event fluctuations of 
the number of charm pairs is different from the Poissonian one. 

From the formal point of view, there is no problem to apply the canonical 
approach to the charm coalescence. 
Let us choose the probability distribution of the event-by-event fluctuations
of the number of $c\bar{c}$ pairs as 
\begin{equation}\label{probCE} 
w_{CE}(N_{c\bar{c}}) = \frac{\gamma_c^2 
Z_{N_{c\bar{c}}}(V,T,\{ \tilde{\lambda} \}) }
{\sum_{N_{c\bar{c}}=0}^{\infty} \gamma_c^{2 N_{c\bar{c}}}  
Z_{N_{c\bar{c}}}(V,T,\{ \tilde{\lambda} \})},
\end{equation}
where
\begin{equation}\label{ZNcc} 
Z_{N_{c\bar{c}}} (V,T,\{ \tilde{\lambda} \})
\equiv \left. Z_{N_{c} N_{\bar{c}}} (V,T,\{ \tilde{\lambda} \})
\right|_{N_{c}=N_{\bar{c}} \equiv N_{c\bar{c}}}.
\end{equation}
In the zero approximation (\ref{Z0}), the sum in the denominator can
be expressed via the modified Bessel function $I_0$:
\begin{eqnarray}\label{ZNccr} 
\sum_{N_{c\bar{c}}=0}^{\infty} \gamma_c^{2 N_{c\bar{c}}} 
Z_{N_{c\bar{c}}} &\approx&
\sum_{N_{c\bar{c}}=0}^{\infty} 
\frac{\left( \gamma_c^{2} \tilde{N}_1 \tilde{N}_{\bar{1}} 
\right)^{N_{c\bar{c}}}}{\left( N_{c\bar{c}} ! \right)^2}  \\
&=& 
I_{0} \left( 2 \gamma_c \sqrt{\tilde{N}_1 \tilde{N}_{\bar{1}}} \right).
\nonumber
\end{eqnarray}

Again, to find the number of charm particles with different charm 
content we have to convolute the probability (\ref{probCE}) with the 
expressions (\ref{N1f}) and (\ref{NHf})--(\ref{N3f}) similarly to (\ref{NkP}). 
The zero approximation results  are expressed via the modified Bessel
functions $I_k$, $k=0,\dots,3$:
\begin{eqnarray} 
\langle N_{1} \rangle_{CE}^{(0)}   &\approx& 
\langle N_{\bar{1}} \rangle_{CE}^{(0)}  \approx
\langle N_{c\bar{c}} \rangle \label{N1CE} \\
  &\approx&   
\gamma_c^{(0)} \sqrt{\tilde{N}_1 \tilde{N}_{\bar{1}}} \,
\frac{I_{1} \left( 2 \gamma_c^{(0)} \sqrt{\tilde{N}_1 \tilde{N}_{\bar{1}}} \right)}
{I_{0} \left( 2 \gamma_c^{(0)} \sqrt{\tilde{N}_1 \tilde{N}_{\bar{1}}} \right)},
\nonumber \\
\langle N_H \rangle_{CE}^{(0)}   &=&   \left( \gamma_c^{(0)}
\right)^2 \tilde{N}_H, \label{NHCE} \\
\langle N_2 \rangle_{CE}^{(0)}   &=&  
\left( \gamma_c^{(0)} \right)^2  \tilde{N}_2 \frac{\tilde{N}_{\bar{1}}}{\tilde{N}_1}
\frac{I_{2} \left( 2 \gamma_c^{(0)} \sqrt{\tilde{N}_1 \tilde{N}_{\bar{1}}} \right)}
{I_{0} \left( 2 \gamma_c^{(0)} \sqrt{\tilde{N}_1 \tilde{N}_{\bar{1}}} \right)},
\label{N2CE}\\ 
\langle N_3 \rangle_{CE}^{(0)}    &=&   
\left( \gamma_c^{(0)} \right)^{ 3}   \tilde{N}_3  \left(  
\sqrt{ \frac{\tilde{N}_{\bar{1}}}{\tilde{N}_1} }
\right)^{   3}  
\frac{I_{3} \left( 2 \gamma_c^{(0)} \sqrt{\tilde{N}_1 \tilde{N}_{\bar{1}}} \right)}
{I_{0} \left( 2 \gamma_c^{(0)} \sqrt{\tilde{N}_1 \tilde{N}_{\bar{1}}} \right)}
 . \nonumber \\
\label{N3CE}
\end{eqnarray}
The formula (\ref{NHCE}) for the hidden charm in CE has the same form as the 
corresponding formula in GCE approach (see the middle part of (\ref{NHG})), 
but the result is different, because the value of $\gamma_c^{(0)}$
is not the same. Indeed, this value 
has to be found at given $N_{c\bar{c}}$ from the transcendental
equation (\ref{N1CE}). Its solution may be quite different from 
(\ref{gamcG}). 

The equation (\ref{N1CE}) can be solved analytically 
in two limiting cases: $\langle N_{c\bar{c}} \rangle \gg 1$ and 
$\langle N_{c\bar{c}} \rangle \ll 1$.
In the first case, the ratios of the Bessel functions tends to $1$:
\begin{equation}
\frac{I_{k} \left( x \right)}
{I_{0} \left( x \right)}
\simeq 1 \mbox{\ \   at \ \ } 
x \gg 1, 
\mbox{ $k=1,2,\dots$}
\end{equation}
and the formulas are reduced to those of the grand canonical 
approach.

In  the second case, $\langle N_{c\bar{c}}  \rangle \ll 1$, the 
Bessel functions can be 
replaced by the leading terms of their Taylor expansions:
\begin{equation}
I_{k} \left( x \right) 
\simeq \frac{x^k}{2^k k!} \mbox{\ \   at \ \ } 
x \ll 1, 
\mbox{ $k=0,2,\dots$}.
\end{equation}
The equation (\ref{N1CE}) then becomes
\begin{equation}
\langle N_{c\bar{c}} \rangle \simeq \left( \gamma_c^{(0)} \right)^2
\tilde{N}_1 \tilde{N}_{\bar{1}}.
\end{equation}
The factor $\gamma_c^{(0)}$ can be easily found:
\begin{equation}
\gamma_c^{(0)} \simeq \sqrt{
\frac{\langle N_{c\bar{c}} \rangle}{\tilde{N}_1 \tilde{N}_{\bar{1}}}
}.
\end{equation}
Then the number of hidden, double and triple charm particles can be
calculated:
\begin{eqnarray}
\langle N_H \rangle_{CE}^{(0)} & \simeq &  \langle N_{c\bar{c}} \rangle   
\frac{\tilde{N}_H}{\tilde{N}_1 \tilde{N}_{\bar{1}}}, \label{NHCEa} \\
\langle N_2 \rangle_{CE}^{(0)} & \simeq & \frac{1}{2} \langle N_{c\bar{c}} \rangle^2
\frac{\tilde{N}_2}{\tilde{N}_1^2}, \label{N2CEa}\\ 
\langle N_3 \rangle_{CE}^{(0)} & \simeq & \frac{1}{6}
\langle N_{c\bar{c}} \rangle^3 \frac{\tilde{N}_3}{\tilde{N}_1^3}. \label{N3CEa}
\end{eqnarray}
The result (\ref{NHCEa}) for the hidden charm coincides with that for 
the Poissonian fluctuation case (\ref{NHP}) in the limit 
$\langle N_{c\bar{c}} \rangle \ll 1$. The reason for this coincidence is 
the following property of the both distributions: 
\begin{equation}
w(n) \ll w(1) \mbox{\ \ at \ \ } \langle N_{c\bar{c}} \rangle \ll 1,
\mbox{\ \ } n=2,3,\dots \ .
\end{equation}
Therefore, the hidden charm hadrons are mostly produced in the 
systems containing 
a single $c\bar{c}$ pair. The probabilities to observe exactly
one $c\bar{c}$ pair becomes approximately equal to each other 
for such distributions, 
if the average number of the pairs is the same and is small:
\begin{equation}
w_{P}(1) \simeq w_{CE}(1) \simeq  \langle N_{c\bar{c}} \rangle
\mbox{\ \ at \ \ } \langle N_{c\bar{c}} \rangle \ll 1.
\end{equation}

In contrast, the double and triple charm particles cannot be produced 
in the system containing only one $c\bar{c}$ pair. Two and three pairs 
at least are needed to produce, respectively, a double and a triple charm 
particle. The probability to have two or three pairs in the system 
are different in the Poissonian and the canonical cases. From this reason, 
the results are essentially different.

Although the Poissonian and canonical probability laws give the same result 
for the hidden charm in two limiting cases:  
$\langle N_{c\bar{c}} \rangle \ll 1$ and $\langle N_{c\bar{c}} \rangle \gg 1$,
they are up to about 10\% different in the intermediate region 
$\langle N_{c\bar{c}} \rangle \sim 1$ \cite{We}.

Which of two approaches, the Poissonian or the Canonical one, 
is appropriate for the description of the charm 
coalescence in a system containing a small number of 
$c\bar{c}$ pairs? 
In fact, the key assumption (\ref{probCE}) of the canonical 
approach relates the fluctuations of the number of charm pairs to
the thermodynamic parameters of the system at chemical freeze-out.
This is in an obvious contradiction  with the basic postulates of the 
statistical coalescence model. Indeed, we have postulated that charm 
quarks are produced
exclusively at the initial stage of the reaction in hard parton collisions. 
Creation and annihilation at later stages are neglected. Therefore, the 
fluctuations of the number of the $c\bar{c}$ pairs cannot have any relation 
to the properties of the system at the thermal stage. From these reasons, 
the canonical approach is not appropriate for the treatment of charm
coalescence. 

The case of strangeness hadronization is essentially different: strange
quarks and antiquarks can be produced at the thermal stage, as far as the 
temperature is larger or comparable to the strange quark mass. Therefore,
in the case of full strangeness thermalization (if not only the momenta, but 
the number of strange quark pairs is thermal), the canonical approach is 
the most appropriate. Still, it should be used with care if 
the full strangeness thermalization is not reached: $\gamma_s \not= 1$.

\section{Summary and Outlook}
\label{Summ}

The production of particles with single, double, triple and 
hidden charm in the framework of the statistical coalescence model
has been considered. 
The grand canonical approach (Section \ref{Grand}) is appropriate for 
systems containing a large number of charm quark-antiquark pairs: 
$\langle N_{c\bar{c}}\rangle \gg 1$. The solution can be found 
analytically (\ref{N1G})--(\ref{N3G}),
if the number of hidden, double and triple charm particles
is small comparing to the total charm. This is the case at presently 
available collision energies. At higher energies, the result can be found 
numerically from the coupled equations (\ref{eq1}) and (\ref{eq2}).

The grand canonical approach cannot be applied to the system containing 
a small number of charm quark-antiquark pairs: 
$\langle N_{c\bar{c}}\rangle \alt 1$. It has been shown that the canonical 
approach is also inappropriate in this case. It is in variance with the basic 
postulates of the statistical coalescence model. As far as 
$c\bar{c}$ pairs are created in mutually independent nucleon-nucleon 
collisions, the fluctuations of $N_{c\bar{c}}$ follows the Poissonian
law. In this case, the result coincides with the grand canonical one 
for the double and triple charm (\ref{N2P}),(\ref{N3P}), but differs 
essentially for the number of the hidden charm (\ref{NHP}). 

The charm coalescence in a part of a large system has been also studied. 
It is sufficient to know the thermal parameters of the subsystem under
consideration to find the number of particles with different charm content. 
The thermal properties of the entire system are not needed. Still one has to
know the total number of the charm pairs in the entire system to calculate
the number of hidden charm particles
in the subsystem. For double and triple charm, 
the information on the number of $c\bar{c}$ pairs in the subsystem suffices.

The obtained derived allow to obtain predictions for double, triple and
hidden charm production in heavy ion collisions at all collision energies.
At very high energies, like  those of the Large Hadron Collider (LHC) in 
CERN (Switzerland), 
the expected number of produced charm quark-antiquarks is large 
(of order of hundreds or even more). Under these conditions, a 
sizable number of double 
and triple charm is expected. 
On the other hand, it would be interesting to see, whether the 
statistical coalescence model works at low energies. The possibility
to study this at the accelerator facility with a very 
high luminosity which is planned to be build in GSI (Germany) is 
worth to be checked. A thermal or nonthermal behavior of heavy quarks 
can tell us much about the properties of the medium created during the heavy
ion collision \cite{thermaliz}.

\begin{acknowledgments}
I acknowledge the financial support of the Deutsche Forshunggemanschaft (DFG),
Germany.
\end{acknowledgments}

\end{document}